\documentclass[aps,showpacs,showkeys,12pt,preprintnumbers]{revtex4}
\usepackage{graphicx}
\usepackage{color}
\usepackage{amsmath}
\usepackage{amssymb}
\usepackage{multirow}
\usepackage{titlesec}
\makeatletter
\renewcommand\paragraph{\@startsection{paragraph}{4}{\z@}%
            {-2.5ex\@plus -1ex \@minus -.25ex}%
            {1.25ex \@plus .25ex}%
            {\normalfont\normalsize\bfseries}}
\makeatother
\def \be  {\begin{equation}}
\def \ee  {\end{equation}}
\def \ee  {\end{equation}}
\def \bea {\begin{eqnarray}}
\def \eea {\end{eqnarray}}

\newcommand{\nn}{\nonumber}

\begin{document}

\title{Hadronization correspondence of Hawking-Unruh radiation from rotating and electrically charged black holes}

\author{Hayam Yassin}
\email{hiam_hussien@women.asu.edu.eg}
\affiliation{Physics Department, Faculty of Women for Arts, Science and Education, Ain Shams University, 11577 Cairo, Egypt}

\author{Eman R. Abo Elyazeed}
\email{eman.reda@women.asu.edu.eg}
\affiliation{Physics Department, Faculty of Women for Arts, Science and Education, Ain Shams University, 11577 Cairo, Egypt}

\author{Raghda E. Break}
\affiliation{Physics Department, Faculty of Women for Arts, Science and Education, Ain Shams University, 11577 Cairo, Egypt}

\author{Amira M. Megahed}
\affiliation{Physics Department, Faculty of Women for Arts, Science and Education, Ain Shams University, 11577 Cairo, Egypt}

\author{Abdel Nasser  Tawfik}
\email{atawfik@nu.edu.eg}
\affiliation{Nile University - Egyptian Center for Theoretical Physics (ECTP), Juhayna Square off 26th-July-Corridor, 12588 Giza, Egypt}

\begin{abstract}
The proposed correspondence between the Hawking-Unruh radiation mechanism in rotating, electrically-charged and electrically-charged-rotating black holes and the hadronization in high-energy physics is assumed. This allows us to determine the well-profound freezeout parameters of the heavy-ion collisions. Furthermore, black holes thermodynamics is found analogical to that of the high-energy collisions. We also introduce a relation expressing the dependence of the angular momentum and the angular velocity deduced from rotating black holes on the chemical potential. The novel phase diagram for rotating, electrically-charged and electrically-charged-rotating black holes are found in an excellent agreement with the phase diagrams drawn for electrically-charged black holes and also with the ones mapped out from the statistical thermal models and the high-energy experiments. Moreover, our estimations for the freezeout conditions $\langle E\rangle/\langle N\rangle$ and $s/T^3$ are in excellent good agreement with the ones determined from the hadronization process, especially at $\mu\leq 0.3$ GeV.
\end{abstract}


\date{\today}

\maketitle


\section{Introduction}
\label{intro}

From various high-energy experiments, such as the ones at the Relativistic Heavy-Ion Collider (RHIC) and the Large Hadron Collider (LHC) \cite{Bjorken:1982qr,Ullrich:2013qwa,Gyulassy:2004zy}, an essential framework for the strongly interacting matter could be constructed \cite{Adcox:2004mh,Back:2004je,Adams:2005dq,Aad:2013xma,Khuntia:2017ite}. Accordingly, it is conjectured that the quark-gluon plasma (QGP) is likely created \cite{Gyulassy:2004zy,Heinz:2015lpa}, i.e. deconfinement phase transition takes place. The created fireball of colorless partons expands and then cools down, rapidly \cite{Tawfik:2014eba}.
The {\it critical} temperature of such hadronization process depends on the net-baryon density of the colliding system or the baryon chemical potential \cite{Ryu:2015vwa,Gale:2012rq,Heinz:2013th,Karpenko:2015xea}. The latter can  phenomenologically be related to both collision energy and critical temperature \cite{Tawfik:2014eba,Grillo:1979wu}. 

From solutions to the general theory of gravity, a classical theory, it is known that the gravitational fields of the black holes (BHs) capture anything approaching the event of horizon. Also, nothing can from escape BHs. But when integrating quantum mechanical effects such as tunneling processes, it turns out that BHs can radiate particles and/or radiations \cite{Hawking:1974rv,Hawking:1974sw,Unruh:1976db,Castorina:2007eb}. These quantum processes are known as Unruh mechanism \cite{Hawking:1974sw,Parikh:1999mf}. The Hawking-Unruh radiation is related to the BHs mass so that heavier BHs radiate more than the lighter ones \cite{Hawking:1974rv,Hawking:1974sw}. There is no constrain limiting/ending such a radiation process \cite{Tawfik:2015kga,Tawfik:2015pqa}.

Such a picture assumes an analogy to the hadronization process in high-energy physics \cite{Chapline:1975tn,Recami:1975px,Sivaram:1975dt,Grillo:1979wu,Parikh:1999mf,Kerner:2006vu,Vanzo:2011wq,Tawfik:2015fda,Tawfik:2016tfe} that the corresponding pair creation and the string breaking could be related to each other \cite{Casher:1978wy,Andersson:1983ia,Becattini:1995if}. With this regard, there is another correspondence to be highlighted, namely the equivalence between BHs metric with a uniform acceleration with the surface gravity and the high-energy system metric, such as Rindler metric, with a uniform acceleration \cite{Castorina:2008gf}.

As introduced in ref. \cite{Isham:1971gm,Salam:1977yr,Castorina:2007eb,Becattini:2008tx}, such a correspondence could be used to find a solution for many still-unsolved puzzles of thermalization and also the freezeout conditions \cite{Castorina:2007eb,Tawfik:2015fda,Tawfik:2016tfe}. Furthermore, the freezeout temperature of the hadronization process was even studied in BHs with a negative cosmological constant \cite{Frassino:2017htb}. It was found that the temperature decreases with the increase in the chemical potential due to the effect of the cosmological constant \cite{Frassino:2017htb}. Prior, similar results was also reported in \cite{Tawfik:2015fda,Tawfik:2016tfe}. 

At vanishing chemical potential, the freezeout conditions $\langle E \rangle / \langle N \rangle=\sqrt{2\, \pi\, \sigma}\simeq 1.09~$GeV and $s/T_f^3=3\, \pi^2\, /4 \simeq 7.4$, at the freezeout temperature $T_f=\sqrt{\sigma/2\, \pi} \simeq 165~$MeV, have been determined \cite{Castorina:2014fna}. It was concluded that the values agree well with the ones obtained from other statistical models \cite{Castorina:2014fna}. This analogy was extended to finite chemical potential and utilized to electrically-charged black holes \cite{Tawfik:2015fda}. An expression for the dependence of the BHs charge on the baryon chemical potential was proposed and then used to calculate the freezeout conditions, which found in good agreement with the heavy-ion collisions, especially at $\mu \leq 0.3$ GeV \cite{Tawfik:2015fda}.

In the present script, we propose to associate the variation of black holes charge with a change in the baryon chemical potential in high-energy collisions, while the variation of the angular momentum of black holes with the change in the angular velocity of the spectators which could be found in offcentral collisions \cite{Castorina:2007eb}. We also propose that the angular velocity can be expressed by the isospin chemical potential, Eqs. (\ref{mu_J}), which is related to the rotating particles in the nuclear collisions, section {\ref{sec:thrm}}. So the temperature of the Hawking-Unruh radiation from all types of black holes is related to the number density and/or the spin related to the principle of collectivity \cite{Castorina:2007eb}.

The present paper is organized as follows. The analogy between the Hawking-Unruh radiation and the hadronization in the high-energy collisions is reviewed in section {\ref{sec:corres}}. The rotating, electrically-charged and electrically-charged-rotating black holes shall be discussed in section {\ref{sec:BHkinds}}. In section {\ref{sec:thrm}}, black holes thermodynamics shall be outlined. Also, the dependence of the angular momentum and the angular velocity of the black holes on the chemical potential is introduced in section {\ref{sec:thrm}}. In section {\ref{sec:frzout}}, the freezeout diagram and both freezeout conditions $\langle E\rangle/\langle N\rangle$ and $s/T^3$ shall be elaborated. The results are discussed in section {\ref{sec:res}}. Section {\ref{sec:concl}} concludes this work.

\section{Analogy between Hawking-Unruh radiation and QCD hadronization}
\label{sec:corres}

The concept of an analogy between the Hawking-Unruh radiation and the QCD hadronization was introduced in various literature \cite{Isham:1971gm,Salam:1977yr,Castorina:2007eb,Becattini:2008tx}. The idea is that both theories are based on the confinement/deconfinement properties. The tunneling quantum process is assumed to form hadrons (hadronization) \cite{Unruh:1976db} and thermal radiation from both high-energy collisions and black holes \cite{Unruh:1976db}. The thermal radiation emitted from BHs is produced as a consequence of the uniform acceleration $\alpha$ of the event horizon \cite{Unruh:1976db,Hawking:1974sw}. Unruh showed that the temperature of this thermal radiation is given as $T=\alpha/2\pi$ \cite{Unruh:1976db}. As discussed, the Hawking-Unruh temperature was found depending on the baryon number density and the angular velocity of the deconfined system \cite{Castorina:2007eb}.

The measurements of high-energy collisions and the statistical thermal models have been used to quantify, as much as possible, the hadronization process and to estimate the freezeout parameters in grandcanonical ensemble \cite{Tawfik:2014eba}. Accordingly, the chemical potential $\mu$ can be estimated as $\mu = n_{\mathrm{B}} \mu_{\mathrm{B}}+n_{\mathrm{S}} \mu_{\mathrm{S}}+n_{\mathrm{I_3}} \mu_{\mathrm{I_3}}$, where $n_{\mathrm{B}}$, $n_{\mathrm{S}}$, and $n_{\mathrm{I_3}}$ are baryon, strangeness, and isospin quantum numbers of the system, respectively \cite{Tawfik:2014dha,Tawfik:2017oyn}. $\mu_{\mathrm{B}}$, $\mu_{\mathrm{S}}$ and $\mu_{\mathrm{I_3}}$ are baryon, strangeness, and isospin chemical potential, respectively. Thus, $\mu_{\mathrm{B}}$ in high-energy collisions could be related to $\mu_{\mathrm{Q}}$ for electrically-charged BHs and $\mu_{\mathrm{I_3}}$ to $\mu_{\mathrm{J}}$ for rotating BHs.

The consideration of adding new BH properties, such as electric charge, in order to extend the chemical potential $\mu$ from vanishing to finite values and accordingly, relate Hawking-Unruh radiation with the hadronization temperature was introduced in ref. \cite{Castorina:2007cm,Tawfik:2015fda}. Also, the dependence of spin or angular momentum in the BH quantum tunneling processes which could then be related to the angular velocity was proposed in ref. \cite{Castorina:2007eb}. In doing this, a new component shall added to the chemical potential $\mu$. This is the isospin chemical potential $\mu_{\mathrm{J}}$ enabling us to express the change in spin of black holes (whether just rotating or electrically-charged-rotating) due to the change in the angular velocity.

\section{Proposed black holes}
\label{sec:BHkinds}

A black hole is formed as a result of the gravitational collapsing of giant stars \cite{Fang:1984me}. The black hole has three basic properties;  mass $M$, charge $Q$, and spin or angular momentum $J$. There are different types of black holes depending on these properties. Also, there are other features such as the event horizon metric and the Hawking-Unruh temperature which could be expressed for the various types of black holes \cite{Castorina:2007eb}. 

The event horizon is formed by the strong gravitational attraction, at which a divergence of Schwarzschild metric. The spacetime line element $ds^2$ reads
\begin{equation}
ds^2 = ( 1 - 2GM/r_{\mathrm{S}})~\!dt^2 - {1\over 1- 2GM/r_{\mathrm{S}}}~\!dr^2,
\label{Schwarz}
\end{equation}
where $t$ is the time. Obviously, $ds^2$ diverges, at the Schwarzschild radius,  $r_{\mathrm{S}} = 2GM$. Where $G=1/2 \sigma = 2.631$ with $\sigma$ is known as the string tension, $\sigma=0.19$ GeV$^2$.

The field strength of the interaction can be obtained from the coefficient
\bea
f(r_{\mathrm{S}}) &=& 1 - 2 \frac{G M}{r_{\mathrm{S}}}.
\eea
The radiation temperature is to be estimated as $f'(r_{\mathrm{S}})/4\pi$ \cite{book_web}
\bea
T &=& \frac{1}{2 \pi} \frac{G M}{r_{\mathrm{S}}^2},
\eea
where $r_{\mathrm{S}}=2GM$. The temperature of Hawking-Unruh radiation is given as
\bea
T_{\mathrm{BH}}(M,0) &=& \frac{1}{8\pi G M},\label{eq:TrS}
\eea
where $T_{BH}(M,0)$ is the Hawking-Unruh radiation temperature from Schwarzschild black hole, i.e. non-rotating and uncharged black hole. 

In the sections that follow, we extend the discussion to cover rotating, electrically-charged and electrically-charged-rotating black holes. We determined $ds^2$, and how $T_{\mathrm{BH}}$ varies.

\subsection{Electrically-charged black holes}

If the black hole has an electric charge $Q$, then the corresponding Coulomb repulsion weakens the gravitational attraction so that the event horizon experiences modifications \cite{Ray:1992}. The Reissner-Nordstr\"om metric (for electrically-charged black holes) can be written as
\begin{equation}
ds^2 = \left( 1 - \frac{2GM}{r_{\mathrm{R}}} + \frac{GQ^2}{r_{\mathrm{R}}^2}\right)dt^2 - \left( 1- \frac{2GM}{r_{\mathrm{R}}} + \frac{GQ^2}{r_{\mathrm{R}}^2}\right)^{-1}dr^2.
\label{RN}
\end{equation}
The divergence leads to smaller Reissner-Nordstr\"om radius
\begin{equation}
r_{\mathrm{R}} = \frac{r_{\mathrm{S}}}{2}\left( 1 + \sqrt{1 - \frac{Q^2}{GM^2}}\right).
\label{RNradius}
\end{equation}
At $Q=0$, $r_{\mathrm{R}}$ reduces to $r_{\mathrm{S}}$. 

The temperature of the Reissner-Nordstr\"om radiation becomes \cite{Fang:1984me,Iso:2006xj,Tawfik:2015fda}
\begin{equation}
T_{\mathrm{BH}}(M,Q) = T_{\mathrm{BH}}(M,0)\left\lbrace4~\sqrt{1 - Q^2/ GM^2}
\over \left[1 + \sqrt{1- Q^2/GM^2}~\!\right]^2\right\rbrace,
\label{T-Q}
\end{equation}
where $T_{\mathrm{BH}}(M,0)$ is the temperature of the Schwarzschild radiation, $
T_{\mathrm{BH}}(M,0)=(8 \pi G M)^{-1}$ \cite{Castorina:2007eb}.

\subsection{Rotating black holes}

If the black hole rotates around an axis, so it will have an angular momentum, which makes the centripetal force opposes the gravitational attraction, which in turn weakens its strength. The resulting Kerr metric for rotating black holes, where $J\neq0$ and $Q=0$, \cite{Castorina:2007eb} is given as
\begin{eqnarray}
ds^2 &=& \left(1 - {2GMr_{\mathrm{K}} \over r_{\mathrm{K}}^2 + a^2 \cos^2 \theta}\right) dt^2 - (r_{\mathrm{K}}^2 + a^2 \cos^2 \theta)~\!d\theta^2 - \left( r_{\mathrm{K}}^2 + a^2 \cos^2 \theta \over r_{\mathrm{K}}^2 - 2GMr_{\mathrm{K}} + a^2\right) ~\!dr^2.
\label{Kerr}
\end{eqnarray}
where $\theta$ is the polar axis and here $\theta = 0$ \cite{Castorina:2007eb}. Also, $J$ is defined as the angular momentum of the black hole, which is obviously given by $J=a M$ with $a$ is the angular momentum parameter. At $a=0$, the metric reduces to the Schwarzschild one. 

For rotating black holes, there are two event horizons; actual (internal surface) and ellipsoid (external surface). The region between these two surfaces is known as the ergosphere. In the present calculations, we consider the actual event horizon, only, which refers to the "outter" confinement \cite{Castorina:2007eb}.  The Kerr radius defines the actual event horizon of the rotating black holes \cite{Castorina:2007eb} 
\begin{equation}
r_{\mathrm{K}} = \frac{r_{\mathrm{S}}}{2}\left( 1 + \sqrt{1 - \frac{a^2}{G^2M^2}}~\!\right)
\label{Kradius}
\end{equation}

The temperature of the Kerr radiation reads
\begin{eqnarray}
T_{\mathrm{BH}}(M,J) &=& \frac{4 G M r^2_{\mathrm{K}} - 2 G M a^2 - 2 G M r^2_{\mathrm{K}}}{4 \pi (r^2_{\mathrm{K}}+a^2)^2} =  T_{\mathrm{BH}}(M,0)~\!\left\lbrace{2 \sqrt{1 - \frac{a^2}{G^2M^2}}
\over \left[1 + \sqrt{1- \frac{a^2}{G^2M^2}}\right] }\right\rbrace,
\label{T-J}
\end{eqnarray}
which obviously reduces to the Schwarzschild one, at $a=0$.

\subsection{Electrically-charged-rotating black holes}

For black holes having both electric charge and spin, known as Kerr-Newman ($Q\neq0$ and $J\neq0$), the corresponding Kerr-Newman metric reads \cite{Xu:2015mna}
\begin{eqnarray}
ds^2 &=& \left( 1 + {Q^2 G - 2GMr_{\mathrm{KN}} \over r_{\mathrm{KN}}^2 + a^2 \cos^2 \theta}\right) dt^2 -  (r_{\mathrm{KN}}^2 + a^2 \cos^2 \theta)~\!d\theta^2 \nn \\
&-& \left( r_{\mathrm{KN}}^2 + a^2 \cos^2 \theta \over r_{\mathrm{KN}}^2 - 2GMr_{\mathrm{KN}} + a^2 + Q^2 G\right) ~\!dr^2.
\label{Kerr-Newman}
\end{eqnarray}
The actual event horizon is given as
\begin{equation}
r_{\mathrm{KN}} =\frac{r_{\mathrm{S}}}{2}\left(1 + \sqrt{1 - \frac{Q^2}{GM^2} - \frac{a^2}{G^2M^2}}\right) ,
\label{KNradius}
\end{equation}
and the temperature of the Kerr-Newman radiation is expressed as
\cite{Fang:1984me,Iso:2006xj}
\begin{eqnarray}
T_{\mathrm{BH}}(M,Q,J) &=& \frac{2 G M r^2_{\mathrm{KN}} - 2 G M a^2 - 2 G Q^2 r_{\mathrm{KN}}}{4 \pi (r^2_{\mathrm{KN}}+a^2)^2} =  \frac{1}{4 \pi} \left\lbrace \frac{2[r_{\mathrm{KN}} - G M]}{r^2_{\mathrm{KN}}+a^2}\right\rbrace \nonumber \\ &=& T_{\mathrm{BH}}(M,0,0) \times \left\lbrace{4 \sqrt{1 - (GQ^2 + a^2)/G^2M^2}
\over \left[1 + \sqrt{1- (GQ^2 +a^2)/G^2M^2}~\!\right]^{~\!2} + a
^2/G^2M^2}\right\rbrace. 
\label{T-QJ}
\end{eqnarray}

\section{Thermodynamics of black holes}
\label{sec:thrm}

For thermodynamics of rotating black holes, we start with the first law of thermodynamics
\begin{equation}
\label{thermodynK}
dE = TdS + \Omega dJ,
\end{equation}
where $S$ is the entropy. $\Omega$ is the angular velocity \cite{Castorina:2007eb} 
\begin{equation}
\Omega = {4\pi \alpha \over S},
\label{pot}
\end{equation}
where the acceleration $\alpha$ can also be given as $\alpha=GM/r^2_{\mathrm{BH}}$. As a result of the analogy with the Hawking-Unruh radiation mechanism, i.e. we relate the temperature to $T_{\mathrm{BH}}(M,J)$ and the angular velocity to $\mu_{\mathrm{J}}$, then the first law  of thermodynamics can be rewritten as 
\begin{equation}
\label{thermodynKN}
dM = T_{\mathrm{BH}}(M,J)\;dS_{\mathrm{K}} + \mu_{\mathrm{J}} dJ,
\end{equation}
where $S_{\mathrm{K}}=\pi (r_{\mathrm{K}}^2 + a^2)/ G$, the entropy of Kerr black holes and $\mu_{\mathrm{J}}$ is the parameter associated with the variation of the angular momentum of the black holes.

Next, we propose an expression relating the chemical potential with rotating BHs (isospin chemical potential) and both angular velocity and angular momentum. 
In general, the rotational kinetic energy of the object can be given as
\bea
E_R &=& Q V, 
\eea
where $E_R= \frac{1}{2} I \Omega^ 2 = \frac{1}{2} j \Omega$, with $I$ is the moment of inertia and $V$ is the potential which will exchanged by the chemical potential $\mu$. Then
\bea
\Omega &=& \frac{2 Q \mu}{j}. 
\eea
Also, we can substitute $Q$ from the relation $\mu =1.4 Q/r$, so that
\bea
\Omega &=& \frac{2 \mu^2 r}{1.4 j}. 
\eea
The dependence of $\mu$ on $\Omega$ can be given
\begin{equation}
\mu_{\mathrm{J}} \sim \sqrt{\frac{\Omega \;a}{0.05\; r_{\mathrm{K}}}},
\label{mu_J}
\end{equation}
where $\mu_{\mathrm{J}}$ and $\Omega$ are in GeV units, while $a$ is dimensionless $r_{\mathrm{K}}$ have the unit of GeV$^{-1}$. Expression (\ref{mu_J}) was proposed due to the relation between the angular momentum and the potential of a particle in the classical theory. From classical theory, we could recall an expression for the rotational kinetic energy of a rotating object in a medium in order to obtain an expression for the relation between $\mu$ and $\Omega$. This expression obviously gives the potential of adding or removing particle from the rotating black hole (isospin chemical potential) and how this depends on the variation of the black hole angular momentum. The chemical potential $\mu$ sums up the various kinds of chemical potentials related to the quantum numbers considered for the system of interest. The correctness of the proposed expression shall be examined in section {\ref{sec:res}}.

Additional to Eq. (\ref{thermodynKN}), the first law of thermodynamics of Kerr-Newman black holes can also reexpressed due to the analogy with the Hawking-Unruh radiation as
\begin{equation}
\label{thermodyn}
dM = T_{\mathrm{BH}}(M,Q,J)\;dS_{\mathrm{KN}} + \mu_{\mathrm{Q}} dQ + \mu_{\mathrm{J}} dJ,
\end{equation}
where $S_{\mathrm{KN}}=\pi (r_{\mathrm{KN}}^2 + a^2)/ G$ and $Q$ are the entropy and the electric charges of the Kerr-Newman black holes, respectively. $\mu_{\mathrm{Q}} = 1.4 Q/r_{\mathrm{R}}$ is the value of the chemical potential related to the variation of the black hole's electric charges \cite{Tawfik:2015fda,Tawfik:2016tfe}. Thus, the chemical potential reads $\mu = \mu_{\mathrm{Q}} + \mu_{\mathrm{J}}$, which expresses the change in the electric charge and the angular momentum of the black holes 
\begin{equation}
\mu = \frac{1.4\;Q}{r_{\mathrm{KN}}} + \sqrt{\frac{\Omega \;a}{0.05\; r_{\mathrm{KN}}}} .
\label{mu_all}
\end{equation}

So far, we conclude that the correspondence between black hole thermodynamics and thermodynamics of Hawking-Unruh radiation can be understood as the variation of black hole's electric charge, which apparently refers to an absorption (or emission) of a particle with a specific baryon chemical potential, while the variation of the black hole angular momentum (lead to changing the angular velocity of BH) indicates an insertion of a new quantum number for these particles known as the "isospin angular momenta", to which we assign an isospin chemical potential.

\section{Freezeout conditions}
\label{sec:frzout}

For the hadron-parton phase transition, we restrict the discussion to the stage of chemical freezeout \cite{Tawfik:2015fda,Tawfik:2013eua,Tawfik:2004ss}. There are different conditions suggested to describe the freezeout phase diagram \cite{Tawfik:2015fda,Tawfik:2013eua,Tawfik:2004ss,Cleymans:1999st,BraunMunzinger:2001mh,Magas:2003wi,
Tawfik:2005qn,Alba:2014eba}. All these conditions are investigated from various statistical thermal models to determine the freezeout parameters $T$ and $\mu$, which in turn have been obtained from measuring particle yields and ratios in various high-energy experiment and then one tries to reproduce these measurements statistical thermal models \cite{Tawfik:2014eba}, where $T$ and $\mu$ are taken as free parameters \cite{Cleymans:1999st,BraunMunzinger:2001mh,Tawfik:2004ss,Tawfik:2013eua,Tawfik:2018ahq}.

We first start with the condition of averaged energy per averaged particle, at vanishing $\mu$ \cite{Castorina:2014fna}
\begin{eqnarray}
\left.\frac{\langle E\rangle}{\langle N\rangle}\right|_{\mu=0}  &=& \sigma\, r_{\mathrm{S}}. 
\end{eqnarray}
At finite $\mu$ (or finite $Q$ and or finite $J$), the string tension of rotating, electrically-charged and electrically-charged-rotating black holes are assumed to take the same phenomenological behaviour \cite{Castorina:2008gf}
\begin{eqnarray}
\sigma(\mu) &\simeq & \sigma(\mu=0)\left[1-\frac{\mu}{\mu_0}\right],
\end{eqnarray}
where $\sigma(\mu=0) = 0.19~$GeV$^2$ and $\mu_0\simeq 1.2~$ GeV is a free parameter \cite{Castorina:2008gf}.
\begin{itemize}
\item At finite $J$, $\mu = \mu_{\mathrm{J}}$, then $r_{\mathrm{S}}$ should be replaced by $r_{\mathrm{K}}$, Eq. (\ref{Kradius}),
which modifies the proposed freezeout condition
\begin{eqnarray}
\left.\frac{\langle E\rangle}{\langle N\rangle}\right|_{\mu\neq0} &=& \sigma(\mu)\, \left\lbrace r_{\mathrm{K}} + a\right\rbrace,
\label{eq:EON_Kerr}
\end{eqnarray}
\item At finite $Q$ and $J$, $\mu = \mu_{\mathrm{Q}} + \mu_{\mathrm{J}}$, then $r_{\mathrm{S}}$ should be replaced by $r_{\mathrm{KN}}$, Eq. (\ref{KNradius}),
and the freezeout condition becomes
\begin{eqnarray}
\left.\frac{\langle E\rangle}{\langle N\rangle}\right|_{\mu\neq0} &=& \sigma(\mu)\, \left\lbrace r_{\mathrm{KN}} + a\right\rbrace.
\label{eq:EON_KerrNew}
\end{eqnarray}
\end{itemize}

Then, we move to the second freezeout condition, namely, the entropy density $s=S/V$ normalized to temperature cubed normalized was proposed to have a constant value in the particle production, vanishing and finite baryon chemical potential \cite{Tawfik:2005qn,Tawfik:2004ss}. The volume $V$ is modelled as $V=4\, \pi\, (r^3 + a^3)/3$, the volume of the black holes is assumed as a sphere. In the Hawking-Unruh radiation from Schwarzschild black hole \cite{Castorina:2014fna}, this condition was estimated as $\left.s/T^3\right|_{\mu =0} = {3}/{8\, G^2\, M\, T^3_{\mathrm{BH}}(M,0)}$ \cite{Tawfik:2015fda}. This condition was also calculated for uncharged and charged black holes and it was concluded that the result agree with the one for the particle production \cite{Castorina:2014fna,Tawfik:2015fda}.

The expressions used to determined $s/T^3$ can be detailed as follows.
\begin{itemize}
\item For rotating black holes, we present that the proposed correspondence of the  Hawking-Unruh radiation and the QCD hadronization is valid, at finite density $\mu = \mu_{\mathrm{J}}$ or (finite $J$)
 \begin{eqnarray}
\left.\frac{s}{T^3}\right|_{J \neq 0} = \left.\frac{s}{T^3}\right|_{\mu =0} \left\lbrace\frac{\left[1+\sqrt{1-\frac{a^2}{G^2\, M^2}}\right]}{2\left[1-\frac{a^2}{G^2\, M^2}\right]^{3/2}}\right\rbrace. \label{eq:sT3Kerr}
 \end{eqnarray}

\item For electrically-charged-rotating black holes, we introduce that the proposed correspondence of the Hawking-Unruh radiation and the QCD hadronization is valid, at finite density $\mu = \mu_{\mathrm{Q}} + \mu_{\mathrm{J}}$ or (finite $Q$ and $J$)
 \begin{eqnarray}
&&\left.\frac{s}{T^3}\right|_{Q \mathrm{and} J \neq 0} = \left.\frac{s}{T^3}\right|_{\mu =0} \times \left\lbrace\frac{\left[\left(1+\sqrt{1-\frac{G Q^2+a^2}{G^2\, M^2}}\right)^2+\frac{a^2}{G^2 M^2}\right]^4}{32\left[1+\sqrt{1-\frac{G Q^2+a^2}{G^2\, M^2}}\right]^3 \left[1-\frac{G Q^2+a^2}{G^2\, M^2}\right]^{3/2}}\right\rbrace.  \label{eq:sT3KerrNew}
 \end{eqnarray}
\end{itemize}

\section{Results and Discussion}
\label{sec:res}

\begin{figure}[htb]
\center
\includegraphics[width=10cm]{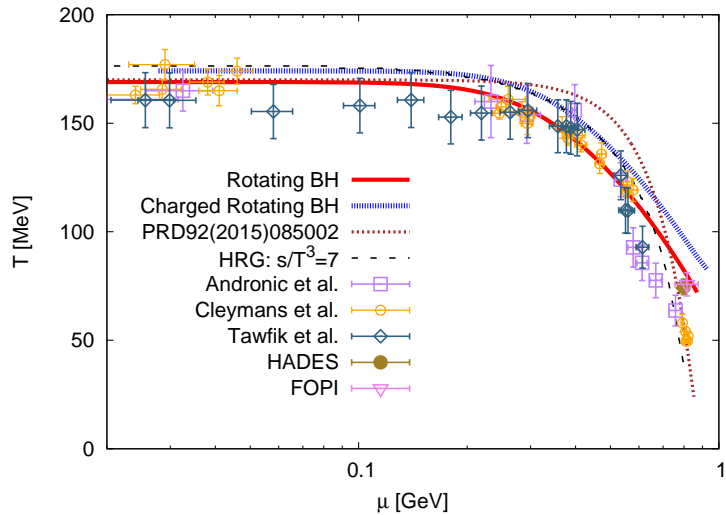}
\caption{(Color online) The freezeout temperature $T$ is depicted in dependence on the chemical potential $\mu$. The calculations from Eqs. (\ref{T-J}, \ref{mu_J}) and Eqs. (\ref{T-QJ}, \ref{mu_all}) are given as solid and double-dotted curves, respectively. The previous results for electrically-charged BHs are represented by the dotted curve, \cite{Tawfik:2015fda}. The dashed curve refers to the HRG results,  at $s/T^3=7$. The symbols depict the freezeout parameters deduced from the particle ratios by using different statistical thermal models: Andronic {\it et al.} \cite{Andronic:2008gu}, Cleymans {\it et al.} \cite{Cleymans:2005xv}, Tawfik {\it et al.} \cite{Tawfik:2018ahq}, and from experimentally measurements for particle ratios: HADES \cite{Agakishiev:2010rs} and FOPI \cite{Lopez:2007aa}.}
\label{fig:1} 
\end{figure}

\begin{figure}[htb]
\center
\includegraphics[width=0.45\textwidth]{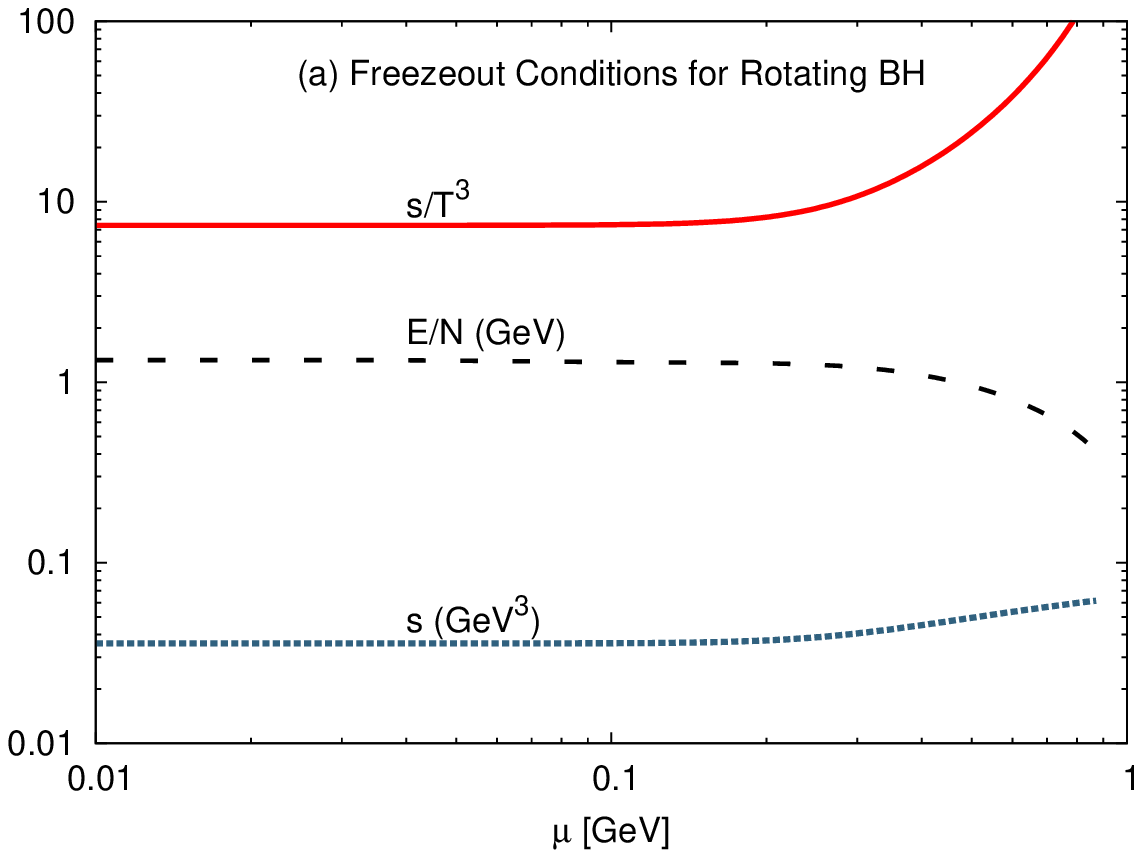}
\includegraphics[width=0.45\textwidth]{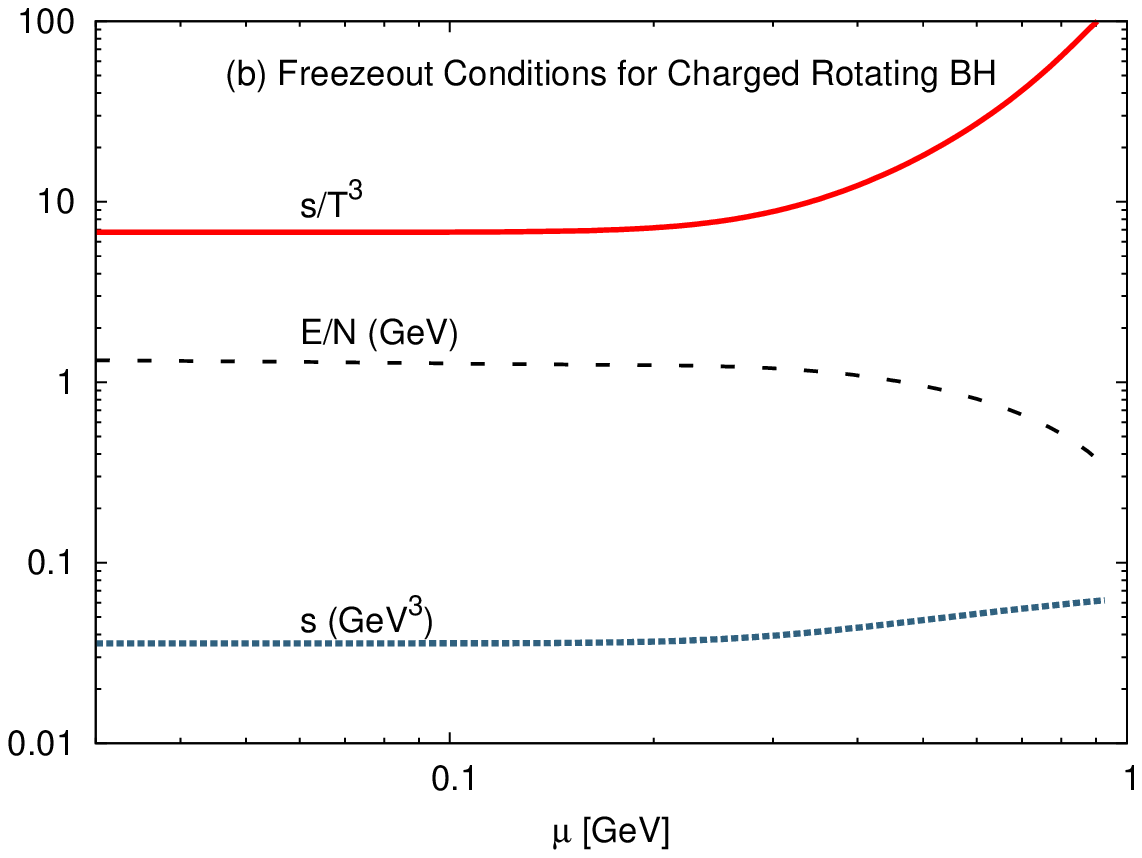}
\caption{(Color online) Left-hand panel shows the freezeout conditions $s/T^3$ (solid curve) and $\langle E\rangle/\langle N\rangle$ (dashed curve) as functions of the chemical potential $\mu$ as calculated for rotating BHs. For a comparison, we also depict the $\mu$-dependence of entropy density $s$ (dotted curve). Right-hand panel shows the same but for electrically-charged-rotating BHs.
}
\label{fig:2}
\end{figure}

Figure {\ref{fig:1}} presents the dependence of the freezeout temperature $T$ on the chemical potential $\mu$; freezeout phase diagram. The solid curve represents the present calculations for the rotating black holes, Eq. (\ref{T-J}), where the variation of $J$ is associated with $\mu=\mu_{\mathrm{J}}$ as given in Eq. (\ref{mu_J}). The double-dotted curve depicts the results for the electrically-charged-rotating black holes, Eq. (\ref{T-QJ}), where the variation of $Q$ and $J$ are associated to $\mu$ as given in Eq. (\ref{mu_all}). Here, we use both the metric for rotating BHs, Eq. (\ref{Kerr}), and the metric for the electrically-charged-rotating BHs, Eq. \ref{Kerr-Newman}, but at $\theta=0$. All present results are compared with the calculations from the electrically-charged black holes (dotted curve) \cite{Tawfik:2015fda}, with the hadron resonance gas (HRG) model at $s/T^3=7$ (dashed line) and with the freezeout parameters (symbols) deduced from the various particle ratios measured and then fitted to the statistical thermal models: Andronic {\it et al.} \cite{Andronic:2008gu}, Cleymans {\it et al.} \cite{Cleymans:2005xv}, Tawfik {\it et al.} \cite{Tawfik:2018ahq}, and from experimental measurements for particle ratios: HADES \cite{Agakishiev:2010rs}, and FOPI \cite{Lopez:2007aa}. 

As a result of the variation in the spin of both types of BHs, the corresponding chemical potential experiences a change and accordingly directly affects the freezeout temperature. We notice that there is an excellent agreement between the present freezeout parameters for both types of black holes \cite{Tawfik:2015fda,Tawfik:2016tfe}. It is obvious that our present results seem improving the calculations reported in ref. \cite{Tawfik:2015fda,Tawfik:2016tfe}, where as hadronization correspondence was also utilized on other types of BHs, as well. It is obvious that taking into consideration the rotation of the black hole besides the black hole electric charge counts for the reasons leading to this improvement. The improvement is apparently determined from the direct comparison with the freezeout parameters deduced from the statistical thermal models, Fig. \ref{fig:1}.

We can now confirm the conclusions drawn in ref. \cite{Tawfik:2015fda,Tawfik:2016tfe} that the freezeout parameters $T$ and $\mu$ can be explained even by the immense gravitational deconfinement as that of the black holes and also that the Hawking-Unruh radiation emitted from rotating, electrically-charged and electrically-charged-rotating black holes finds a correspondence in the particle production. We also conclude that the degrees of freedom measured by constant $s/T^3$, for instance, seem to remain conserved during the black hole radiation process, which is accompanied by a mass reduction. We mean that the charge and the angular momentum of BH are conjectured not to affect the reduction of the mass when the radiation is emitted from BH. With decreasing $T$, the BH mass increases preventing BH from the entire dissolving.  But when $J\neq 0$ and/or while $Q \neq 0$, the rotational force and the Coulomb repulsion oppose the gravitational attraction. In the case that these forces become able to overcome the gravitational attraction force, BH dissolves although the BH temperature decreases.

Figure {\ref{fig:2}} depicts the freezeout conditions $s/T^3$ (solid curve) and $\langle E\rangle/\langle N\rangle$ (dashed curve) as functions of the chemical potential $\mu$, Eqs. (\ref{eq:EON_Kerr}, \ref{eq:sT3Kerr}), as deduced from rotating BHs, left-hand panel. For the seek of interpretation, we also draw the entropy density $s$ (dotted curve) as a function of the chemical potential $\mu$. The same is also drawn in the right-hand panel, but here for electrically-charged-rotating black holes using, Eqs. (\ref{eq:EON_KerrNew} and \ref{eq:sT3KerrNew}).

In both cases, we notice that the $s/T^3$ of resulting BHs remains constant $\sim 7$, at $\mu \leq 0.3~$GeV, which agrees well with the value conducted from the particle production in the various high-energy experiments \cite{Tawfik:2005qn,Tawfik:2004ss}. At $\mu > 0.3~$GeV, we find that the entropy density normalized to $T^3$ increases with the increase in $\mu$, while the entropy density $s$ is nearly independent on the change in spin and/or electric charge of the black holes or $\mu$. Thus, the increase in $s/T^3$, at $\mu > 0.3~$GeV, can be understood due to the decrease in the denominator with respect to the numerator or the increase in the fraction within the bracket of the denominator of Eqs. (\ref{eq:sT3Kerr}, \ref{eq:sT3KerrNew}). This can be fulfilled with increasing $Q$ and $J$. The results obtained for $s/T^3$ apparently agree well with the electrically-charged black holes \cite{Tawfik:2015fda}. In the present study, we also have taken into consideration the same possibilities proposed in ref. \cite{Tawfik:2015fda} for the rotating and the electrically-charged-rotating black holes. The mass of black holes which describes well the correspondence between the Hawking-Unruh radiation in both types of black holes and the hadronization in the high-energy experiments is assumed as $1.3606$GeV. In our calculations for electrically-charged-rotating black hole, we have fixed the charge of the black hole to $Q=0.15~$GeV.

The other freezeout condition, namely $\langle E\rangle/\langle N\rangle$, also  remains constant, $\simeq 1.35$GeV, at $\mu \leq 0.3$ GeV. At higher $\mu$, the average energy per particle seems slightly decreasing as a result of the variation of the spin and electric charge of the black holes. This result agrees well with the value proposed in ref. \cite{Cleymans:1999st,Tawfik:2014dha,Tawfik:2016jzk} and also was confirmed in ref. \cite{Tawfik:2015fda,Tawfik:2016tfe}.

\section{Conclusions}
\label{sec:concl}

The correspondence between the hadronization process in high energy physics and the black holes are examined. Precisely, we have studied this analogy for two types of black holes, rotating and electrically-charged-rotating black holes. We have proposed an expression for the dependence of the angular momentum and the angular velocity on the chemical potential. Phase diagram for rotating and electrically-charged-rotating black holes are found to be in an excellent agreement with the ones deduced from the electrically charged black holes and from the different statistical thermal models.

We have calculated the two freezeout conditions, $\langle E\rangle/\langle N\rangle$ for average energy per particle and $s/T^3$ for entropy density ($s$) normalized to cubic freezeout temperature in both types of studied black holes. Both freezeout conditions are in excellent agreement with the ones in the hadronization process till $\mu \leq 0.3$ GeV. At higher values of $mu$, the value of $\langle E\rangle/\langle N\rangle$ was found decreasing below $1.01$ GeV, because this quantity entirely depends on the change of $\mu$ which is related to the change in the angular momentum and the charge of the black holes.

For $s/T^3$, with the further increase of $\mu$ more than $0.3$ GeV, we have found a rapid increase even beyond $7$ as a result of the freezeout temperature because at that values of $\mu$ the entropy density $s$ remains nearly constant. Further increase in $\mu$ (or $J$, $Q$) enhances $s$. Also, it was noticed that this observation agrees well with the value obtained from the electrically-charged black holes which in turn means that the analogy between the hadronization process in high-energy physics and the Hawking-Unruh radiation from all types of black holes can be used to estimate the values of the freezeout conditions, at least partially. In a future work, we plan to use the analogy between QCD and the Hawking-Unruh radiation process in order to study the charm production in both high-energy physics and Hawking-Unruh radiation radiation.

\bibliographystyle{aip}
\bibliography{mybibfile}

\end{document}